\documentclass[prb,twocolumn,aps,amssymb,amsmath,superscriptaddress,showpacs]{revtex4-1}

\usepackage{bm}
\usepackage{graphicx}
\usepackage{color}
\usepackage{amsfonts}
\usepackage{textcomp} % text companion fonts
\usepackage{microtype} % makes pdf looks better
\usepackage{epstopdf}
\usepackage{amsmath,mathtools}
\usepackage{hyperref}
\hypersetup{colorlinks=true}
\usepackage[all]{hypcap} % let hyperlinks correctly point to figures
\graphicspath{{fig/}}

\begin{document}

\title{Quench Dynamics Across Topological Quantum Phase Transitions}

\author{Shiuan-Fan Liou}
\affiliation{National High Magnetic Field Laboratory and Physics Department, Florida State University, Tallahassee, FL 32306, USA}
\author{Kun Yang}
\affiliation{National High Magnetic Field Laboratory and Physics Department, Florida State University, Tallahassee, FL 32306, USA}

\begin{abstract}
We study the dynamics of systems quenched through topological quantum phase transitions and investigate the behavior of the bulk and edge excitations with various quench rates. Specifically, we consider the Haldane model and checkerboard model in slow quench processes with distinct band-touching structures leading to topology changes. The generation of bulk excitations is found to obey the power-law relation Kibble-Zurek and Landau-Zener theories predict. However, an anti-Kibble-Zurek behavior is observed in the edge excitations. The mechanism of excitation generation on edge states is revealed, which explains the anti-Kibble-Zurek behavior.

\end{abstract}

\date{\today}

\maketitle

\section{Introduction}
The nonequilibrium dynamics of systems undergoing phase transitions is an important subject in statistical physics. In particular, quench dynamics through both classical
and quantum second-order phase transitions involving symmetry breaking has been of great interest. The Kibble-Zurek (KZ) mechanism, a theory originally developed in the study of the formation of topological defects in the early universe~\cite{JphysicsA9.1387, physRep67.183}, was applied to quench dynamics near symmetry-breaking second-order phase transitions~\cite{physRep317.505,ZUREK1996177, PhysRevB.86.064304} and provided a fairly accurate prediction of a power-law relation between the topological defect density and the quench rate\cite{PhysRevLett.95.105701}. In the meantime, Landau-Zener (LZ) theory~\cite{zener_1932} describing a two-level transition was also applied to the study of the dynamics of quantum phase transitions where applicable, yielding results~\cite{PhysRevLett.95.035701,PhysRevLett.95.245701} consistent with the KZ theory. Some systems under inhomogeneous quench~\cite{1367-2630-12-5-055007} and non-linear quench~\cite{PhysRevLett.101.016806,PhysRevLett.101.076801} out of
the scope of KZ mechanism have been investigated as well.

In contrast, non-equilibrium dynamics across topological phase transitions that do not involve symmetry breaking has been studied much less. Topological
phases of matter are of tremendous importance and current interest~\cite{RevModPhys.89.040502}.
Some fundamental questions about the dynamics of topological phase transitions naturally arise. First, since there is no symmetry breaking in such phase
transitions, can KZ theory effectively describe their dynamics? Second, what is the mechanism of topological defect generation, and how is it related to the symmetry-breaking case?
Among many unique properties of topological states, of particular importance is the presence of robust edge states that often give rise to dissipationless and quantized transport; they have been proposed to be the building blocks of electronic devices with low or even zero dissipation~\cite{PhysRevLett.101.146802, PhysRevLett.115.057206}. Operating such devices often involves switching between topologically trivial (the insulating, or off) states and nontrivial (conducting, or on) states. Thus, understanding quench dynamics across topological phase transitions, especially its impact on edge states, is of both fundamental and practical importance. While there exist some reports on such studies~\cite{PhysRevLett.102.135702,NewJPhys12.055014,PhysRevB.92.035117,Patel2013, PhysRevLett.115.236403,PhysRevB.94.155104, PhysRevLett.118.185701,arxiv1709.01046,PhysRevB.88.104511, PhysRevLett.113.076403, PhysRevA.92.053620, 0953-8984-25-40-404214, PhysRevE.90.032138}, a comprehensive understanding has yet to emerge, especially when edge states are involved.

In this work, we study quench dynamics across topological quantum phase transitions (TQPTs) in the simplest setting of free-fermion systems. Since in the ultracold-atom field the Haldane model can be realized through optical traps~\cite{nature_exp}, which enables people to explore and access the phase diagram of the Haldane model, some research about the sudden quench dynamics through various topological phase transitions in the Haldane model has been demonstrated in ultracold-atom experiments~\cite{arxiv1709.01046}. We believe the slow quench process of the Haldane model is feasible because sudden quench is just a limit of a general quench process. Thus, here we specifically consider two models: the Haldane model~\cite{PhysRevLett.61.2015} and the checkerboard model~\cite{PhysRevLett.103.046811,PhysRevLett.106.236803} with linear and quadratic band dispersions at the gap-closing points (where the TQPTs occur) respectively. The TQPTs result in changes in the band Chern numbers by 1 and 2 respectively, and the appearance of edges states in the former and the reverse of edge-state chirality in the latter. We numerically follow the time evolutions of the systems (initially in the ground states) under quenches that move the systems across the phase boundaries and monitor the generation of excitations both in the bulk and at the edge. We argue the bulk excitations are analogous to the topological defects in the case of symmetry-breaking phase transitions, and obtain results consistent with the
prediction of KZ and LZ theories.The appearance of edge excitations, on the other hand, is unique to TQPTs and has no analog in symmetry-breaking phase transitions. A particularly interesting finding that we report here is an anti-KZ behavior; namely, the number of edge excitations depends on quench rate non-monotonically. We will provide an explanation of this counter-intuitive result.

The rest of this paper is organized as follows: in Sec.~\ref{model}, we introduce the two models, the Haldane model and checkerboard model, studied in this work. In Sec.~\ref{theory}, we briefly describe the concepts of Kibble-Zurek theory and Landau-Zener theory, derive the power-law relation between the (bulk) excitation density and the quench rate based on each theory, and use the results for our models. Before showing the results, in Sec.~\ref{dynamics} we describe the numerical methods of studying slow quench problems with edges and calculating bulk and edge excitations. We show our results and provide some discussions in Sec.~\ref{result}, in which the discussion of and a comparison with theoretical predictions of the bulk excitations are given in Sec.~\ref{bulk} and the discussion of edge excitation and the mechanism of the excitation generation are included in Sec.~\ref{edge}. In Sec.~\ref{conclusion} we end the paper by offering some concluding remarks.

\section{Models}\label{model}

In this section we introduce the models we study and their phase diagrams.

\subsection{Haldane model}
The Haldane model describes spinless fermions hopping on a honeycomb lattice with a real nearest-neighbor (NN) hopping, a complex next-nearest-neighbor (NNN) hopping, and an
energy offset with a sign difference on the two sublattices. The Hamiltonian can be written as
\begin{equation}
\begin{aligned}
H = &-\sum\limits_{\langle i, j \rangle} \left( C _{A, i} ^{\dagger} C _{B, j} \, + \mbox{H.c.} \right) \\
 &+ \eta \sum\limits_{\langle\langle i, j \rangle\rangle} \left[ \, e ^{i v _{ij} \phi} \left( C _{A, i}
^{\dagger} C _{A, j} - C _{B, i} ^{\dagger} C _{B, j} \right) \, + \mbox{H.c.} \right] \\[0.2cm]
 & +\, m \sum\limits_{i} \left( C _{A, i} ^{\dagger} C _{A, i} - C _{B, i} ^{\dagger} C _{B, i} \right),
\end{aligned}
\end{equation}
where $C _{\sigma, i} ^{\dagger} \left( C _{\sigma, i}  \right)$ is the fermion creation (annihilation) operator, $\langle i, j \rangle$ and $\langle \langle i, j \rangle
\rangle$ represent summing over the NNs and NNNs, respectively, and $A$ and $B$ label the two sublattices. $v _{ij} \equiv \hat{z} \cdot
\left( \hat{d} _{j} \times \hat{d} _{i} \right)$ with $\hat{z}$ being the unit vector perpendicular to the two-dimensional (2D) plane and $\lbrace \hat{d} _{i} \rbrace$ being the unit vector along the bond
connecting two nearest sites, as shown in Fig.~\ref{hdphdg}(a). The complex hopping with a phase $e ^{i v _{ij} \phi}$ in the second term due to the staggered magnetic field
breaks time-reversal symmetry, and the last term breaks spatial-inversion symmetry.

As shown in the phase diagram~\cite{PhysRevLett.61.2015} [Fig.~\ref{hdphdg}(c)], we can access different topological phases by tuning parameters $m$, $\eta$, and $\phi$. In this
study, we quench the system from the topologically trivial phase with Chern number $C = 0$ to the topologically non-trivial phase with $C = -1$ along the path (vertical dashed arrow) in
Fig.~\ref{hdphdg}(c), setting $\eta = \frac{1}{3}$ and $\phi = \frac{\pi}{2}$ ($|\eta| \leq \frac{1}{3}$ prevents band overlap)~\cite{PhysRevLett.61.2015}, by
varying $m$ from 3 to 0 linearly with time, namely, $m = 3 - \frac{t}{\tau}$, with $\frac{1}{\tau}$ being the quench rate. When $m = \sqrt{3}$, the system reaches the phase boundary,
and the energy gap in the bulk closes at the high-symmetry point ($K$ or $K ^{\prime}$), resulting in linear dispersion.
Figures~\ref{dispersion}(a) to \ref{dispersion}(c) show the dispersions of $H \left( t \right)$ of the Haldane model with zigzag edges in the $x$ direction at the initial time, critical time, and final time, corresponding to no edge states,
band touching, and edge states popping up, respectively.

\subsection{Checkerboard model}
\begin{figure}[!t]
\centerline{\includegraphics[width=9cm]{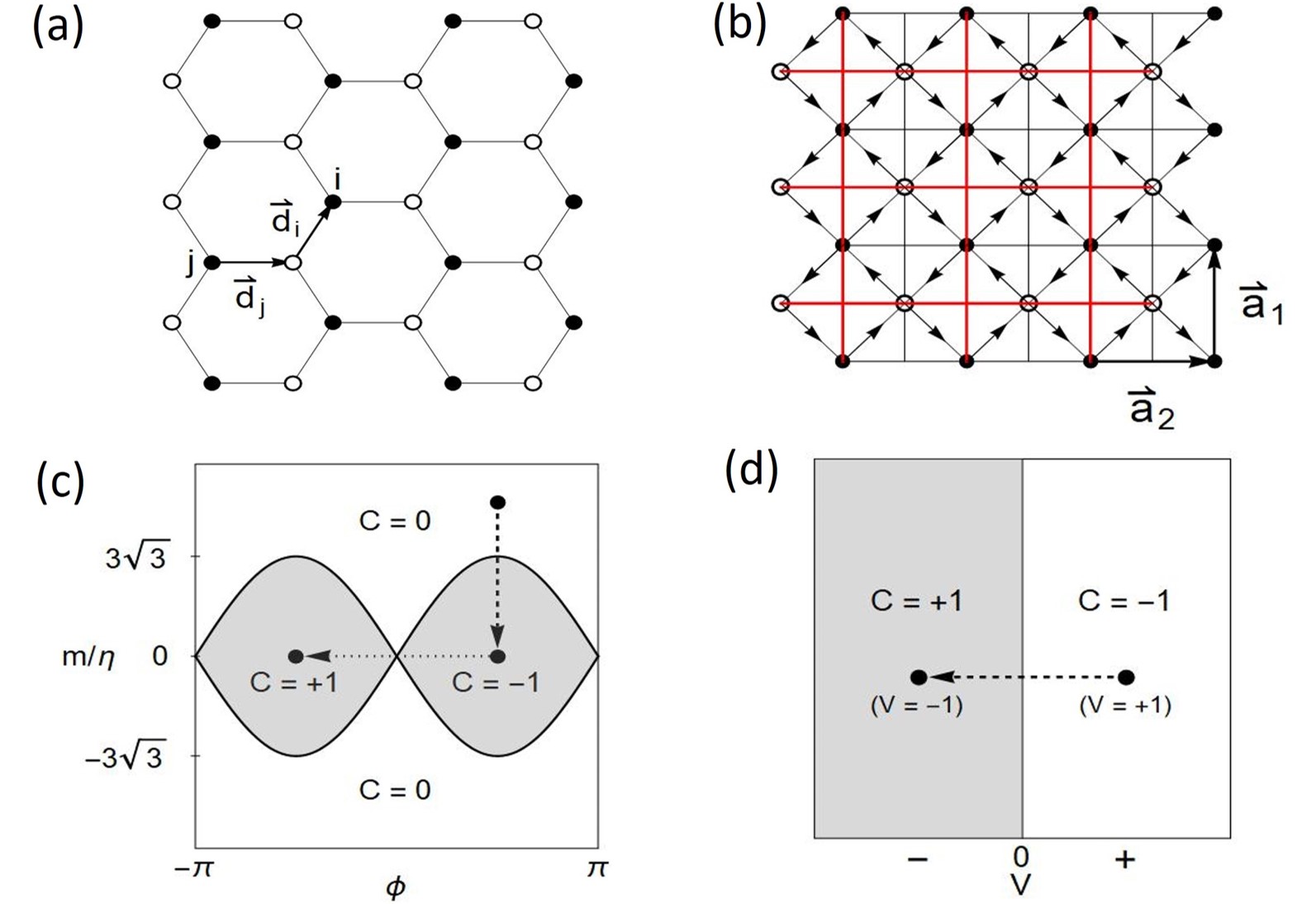}}
\caption{(a) Haldane model on the honeycomb lattice. (b) Checkerboard model. The NNN hoppings along the red and the black
cross lines have hopping strengths $\xi ^{\prime}$ and $\xi ^{\prime \prime}$; the imaginary strength of the NN hoppings following the directions of the arrows is positive, and it is negative otherwise. (c) Phase diagram of the Haldane model. (d) Phase diagram of the checkerboard model with only one relevant parameter, $V$.}
\label{hdphdg}
\end{figure}

Motivated by Refs.~\cite{PhysRevLett.103.046811,PhysRevLett.106.236803}, we consider a simplified checkerboard model with two sublattices ($A$ and $B$) with real and isotropic NN hoppings, two kinds of nontrivial NNN hoppings, and a complex NN hopping which leads to a quantum anomalous Hall (QAH) phase. The Hamiltonian is given by
\begin{equation}
\begin{aligned}
H &=-\sum\limits_{\vec{r}, \vec{\delta}} \, C _{A, \vec{r}} ^{\dagger} \, C _{B, \vec{r} + \vec{\delta}} \, + i V \sum\limits_{\vec{\delta}} D_{\delta} \, C _{A, \vec{r}} ^{\dagger} \, C _{B, \vec{r} + \vec{\delta}} \\[0.2cm]
 & - \xi ^{\prime} \sum\limits_{\vec{r}} \left( C _{A, \vec{r}} ^{\dagger} C _{A, \vec{r} \pm \vec{a} _{1}} + C _{B, \vec{r}} ^{\dagger}  C _{B, \vec{r} \pm \vec{a} _{2}}
 \right) \\[0.2cm]
  &- \xi ^{\prime \prime} \sum\limits_{\vec{r}} \left( C _{A, \vec{r}} ^{\dagger} \, C_{A, \vec{r} \pm \vec{a} _{2}} + C _{B, \vec{r}} ^{\dagger} \, C _{B, \vec{r} \pm \vec{a}
  _{1}} \right) + \mbox{H.c.}, \\[0.2cm]
\end{aligned}
\end{equation}
where $\vec{a} _{1} = \pm \left( 0, a \right)$, and $\vec{a} _{2} = \pm \left( a, 0 \right)$, $D _{\vec{\delta}} = +1$ if $\vec{\delta} = \pm \left(a/2, a/2 \right)$ and $D
_{\vec{\delta}} = -1$ if $\vec{\delta} = \pm \left(a/2, -a/2 \right)$ with $a/\sqrt{2}$ being the lattice spacing. The first term represents the isotropic NN hopping. The second term is a purely imaginary NN hopping which is positive if hopping along arrows in Fig.~\ref{hdphdg}(b)
and is negative otherwise, inducing the QAH phase with nontrivial topology $C = \pm 1$. The last two terms correspond to the NNN hoppings along the red and black cross lines in Fig.~\ref{hdphdg}(b) with strengths $\xi ^{\prime}$ and $\xi ^{\prime \prime}$, respectively, which break $C _{4}$ symmetry if $\xi ^{\prime} \neq \xi ^{\prime \prime}$.

\begin{figure}[!t]
\centerline{\includegraphics[width=9cm]{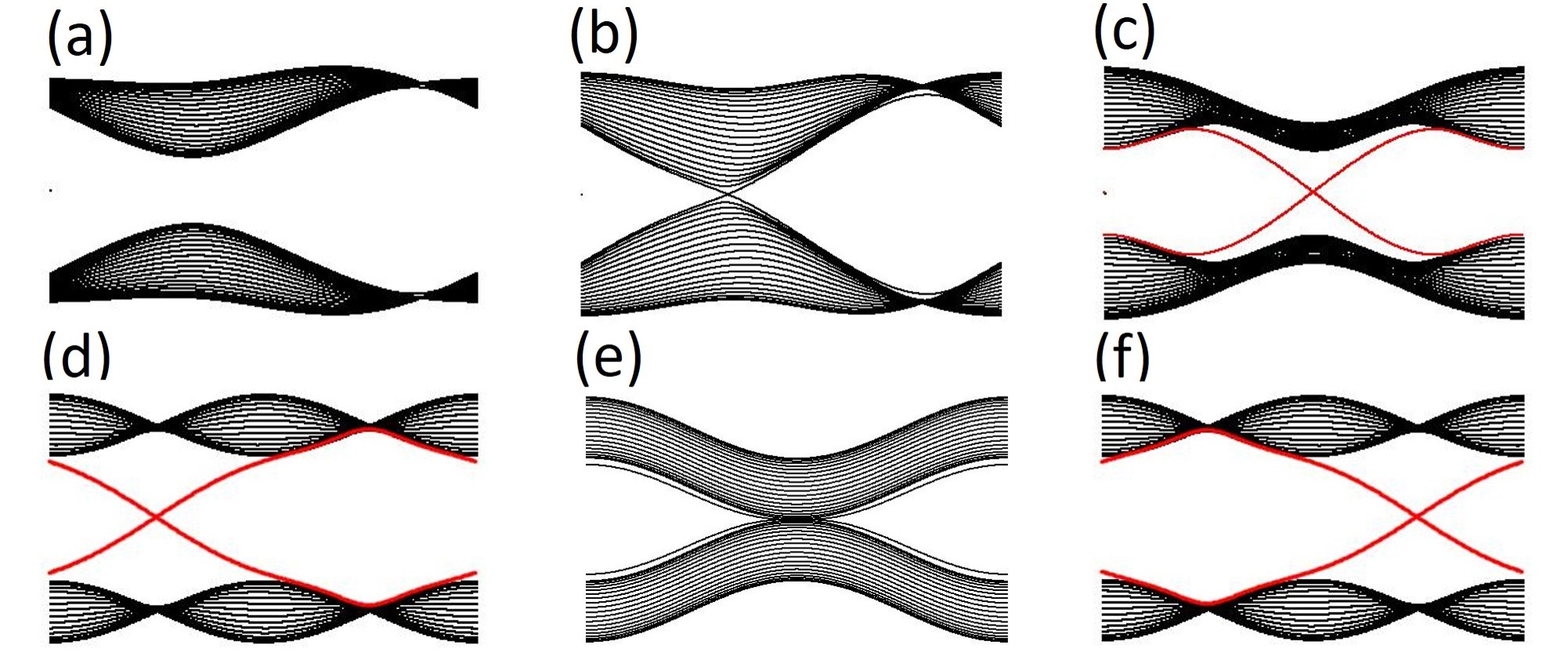}}
\caption{(a)-(c) Dispersions of the Haldane model with the zigzag edge at the initial time ($C = 0$), at the critical time when $m = \sqrt{3}$ with gap closing (linear dispersion), and at the final time ($C = 1$) with edge states (red curves). (d)-(f) Dispersions of the checkerboard model at the initial time ($C = -1$) with edge states on the
left, at the critical time when $V = 0$ with gap closing (quadratic dispersion), and at the final time ($C = +1$) with edge states on the right. Edge states are the red curves.}
\label{dispersion}
\end{figure}

A phase diagram of the checkerboard model with only one relevant parameter, $V$, is provided in Fig.~\ref{hdphdg}(d). In this model, we change the complex hopping strength $V$ from 1
to -1 linearly, $V = 1 - \frac{t}{\tau}$, so that the Chern number $C$ changes sign (from -1 to 1), indicated by the dashed arrow in Fig.~\ref{hdphdg}(d) with $\xi ^{\prime} = - \xi ^{\prime \prime} = 0.5$. At $V = 0$, the gap in the bulk closes at $\left( k_{x}, k _{y} \right) = \left( \pi , \pi \right)$ with
a quadratic dispersion, as shown in Fig.~\ref{dispersion}(e). During the quench process, the dispersions of $H \left( t \right)$ with edges in the $x$ direction at the initial, critical, and final times are shown in Figs.~\ref{dispersion}(d) to \ref{dispersion}(f). Since the system is quenched from one topologically nontrivial phase to another, edge states always exist during the whole process, but their chirality is reversed. Note that one can also change the Chern number by 2 in the Haldane model following the horizontal dotted path in Fig.~\ref{hdphdg}(c). However, the gap closing will occur at two Dirac points, $K$ and $K ^{\prime}$, instead of one, resulting in critical behavior distinct from the case in the checkerboard model but similar to what happens when the gap closes at a single Dirac point, as discussed earlier.

\section{Theories}\label{theory}

In preparation for later comparisons, in this section we review the basics of KZ and LZ theories and, in particular, discuss the relevance of the KZ theory to the topological phase transitions we study.

\subsection{Kibble-Zurek theory}
\begin{figure}[!t]
\centerline{\includegraphics[width=5cm]{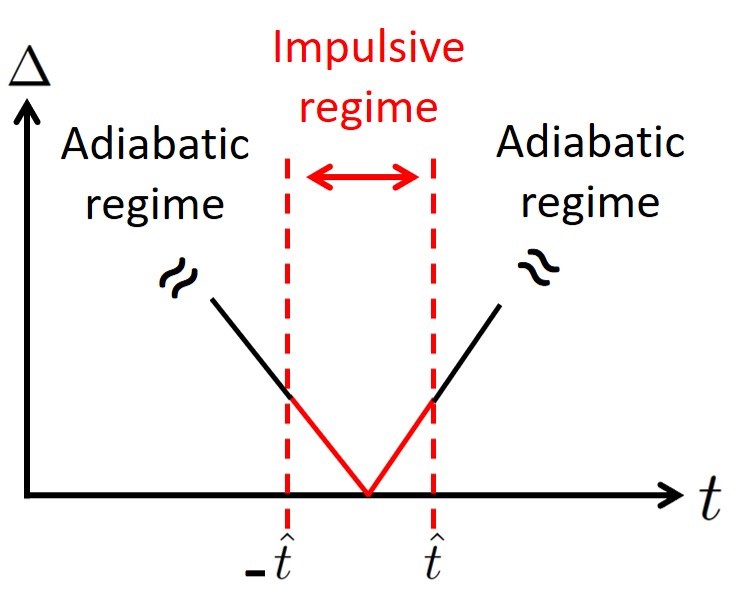}}
\caption{Schematic diagram to illustrate the Kibble-Zurek theory. The energy gap of two states $\Delta$ is a function of time $t$. Within $t = -\hat{t}$ to $\hat{t}$, the system is in the impulsive regime in which the relaxation time is comparable to or much greater than the quench timescale, so that the system evolves diabatically.}
\label{kz_theory}
\end{figure}

For a quench process which involves a second-order phase transition from a high-symmetry phase to a broken-symmetry phase with gap closing at the critical point, the system can no longer evolve adiabatically
due to the unreachable relaxation time which is inversely proportional to the energy gap. KZ theory separates such a quench process into two regimes: the adiabatic regime, and impulsive regime as shown in Fig.~\ref{kz_theory}. When the system enters the impulsive (diabatic) regime, the information of the system will be frozen because the relaxation time and the timescale of the quench are comparable. In such a quench process, (bulk) excitations (defects) are inevitable.

In the following, we derive the relation between the (bulk) defect density and the quench rate based on a linear quench assumption. For a second-order phase transition, the energy gap $\Delta$ has a power-law relation with the quench parameter $\mu \left( t \right)$ with the critical point $\mu _{c}$ where the gap closes: \\
\begin{equation}\label{gap}
\Delta \sim \, \lvert \frac{\mu \left( t \right)}{\mu _{c}} \, - \, 1 \rvert ^{z \nu},
\end{equation}

\noindent where $z$ is the dynamic critical exponent and $\nu$ is the correlation length critical exponent. Moreover, the correlation length $\xi$ has a power-law relation with the quench parameter as well and will blow up for infinite systems or be comparable to the system size for finite systems at the critical point: \\
\begin{equation}\label{correlation}
\xi \sim \, \lvert \frac{\mu \left( t \right)}{\mu _{c}} \, - \, 1 \rvert ^{- \nu}.
\end{equation}

\noindent Assume that the quench process starts at $t = -\infty$ and terminates at $t = \infty$ and the gap closes at $t = 0$. According to the linear quench assumption, we can define $\epsilon \left( t \right) \equiv \frac{\mu \left( t \right)}{\mu _{c}} - 1 \, \sim \, \frac{|t|}{\tau}$, with $\frac{1}{\tau}$ being the quench rate. While the system enters the impulsive regime at time $- \hat{t}$, as shown in Fig.~\ref{kz_theory}, the relaxation time $\eta \left( -\hat{t} \right)$ and the quench time scale $|\hat{t}|$ are comparable, namely,  \\
\begin{equation}\label{eta}
\eta \left( -\hat{t} \right) = \frac{\hbar}{\Delta} \, \sim \, \hat{t}.
\end{equation}

\noindent From Eq.~(\ref{eta}) we can express $\hat{t}$ in terms of the quench rate $\frac{1}{\tau}$. From a high-symmetry phase to a broken-symmetry phase, the excitation size could be estimated through the correlation length as $\xi ^{d}$ in $d$-dimensional space. Combining Eq.~(\ref{gap})-(\ref{eta}), therefore, we get the power-law relation of the excitation density $n _{topo}$, which is inversely proportional to the size of an excitation, and the quench rate \\
\begin{equation}\label{power_law_kz}
n _{topo} \, \sim \, \tau ^{\frac{-d \nu}{1 + z \nu}}.
\end{equation}

\noindent The power-law relation in Eq.~(\ref{power_law_kz}) has been shown to have very good agreement with the numerical result from studying the phase transitions associated with symmetry breaking\cite{PhysRevLett.95.105701}.   

The topological phase transitions in which we are interested in this work are second-order phase transitions. Since there is no symmetry breaking in such transitions, it may not be obvious that the KZ theory is relevant here at first glance. Those topological phases here are characterized by Chern numbers and have one-to-one correspondences to (integer) quantum Hall states. As reviewed in Ref. \cite{zhang1992}, quantum Hall states can be mapped onto superfluid states, in which composite bosons made of electrons and an appropriate amount of flux condense and develop (quasi-)long-range order. They are thus analogous to broken-symmetry states. In particular, excitations (particles in the conduction bands and holes in the valence bands) in quantum Hall states are analogues to vortices and antivortices in the superfluid in this mapping, which are the topological defects of the ordered phase. In retrospect this is rather natural as the topological characterization is specific to the ground state; any excitation on top of the ground state causes a deviation from, say, the perfectly quantized Hall conductance of the ground state and is clearly a topological defect. Thus, the excitations induced by the quenches through topological phase transitions are analogs to the those in ordered phases. This allows us to use the results of KZ theory [Eq.~\ref{kz_theory}] in the Haldane and checkerboard models, which we turn to now. 

Due to linear quench, the gap has a linear dependence on the quench parameter in both the Haldane and checkerboard models, giving $z \nu = 1$ (according to $\Delta \sim |\mu - \mu _{c}| ^{z \nu}$, where $\Delta$ is the gap and $\mu$ is the quench parameter). For the Haldane model, which has $d = 2$, $z = 1$ (linear dispersion), and $\nu = 1$, the predicted power $\alpha$ is 1 from KZ theory ($n _{topo} \sim \tau ^{-\alpha}$); for the checkerboard model, with $d = 2$, $z = 2$ (quadratic dispersion), and $\nu = \frac{1}{2}$, it gives $\alpha = 0.5.$

\subsection{Landau-Zener theory}
LZ theory describes the dynamics of a two-level system with a time-dependent Hamiltonian in which the energy gap of the two states varies linearly with time. Note that LZ theory can be applied to such a two-level system regardless of the topology (trivial or non-trivial) of the states. Suppose that the Hamiltonian evolves with time $t$, from $ t = - \infty$ to $t = + \infty$ in which the gap of the two states has a minimum at $t = 0$ (one can always shift the time such that the gap minimum occurs at $t = 0$), and that initially (at $t = -\infty$), one of the states is occupied and the other is empty. The Hamiltonian with the basis $\left(\psi _{+} , \psi _{-} \right)$ can be written as \\
\begin{equation}
H = \, \begin{pmatrix}
\epsilon _{1}  &  \epsilon _{12} \\
\epsilon _{21} &  \epsilon _{2} \\
\end{pmatrix},
\end{equation}

\noindent where $\psi _{+}$ and $\psi _{-}$ represent the two states with $\psi _{+} = \begin{pmatrix} 1  \\  0 \end{pmatrix}$, which is occupied, and $\psi _{-} = \begin{pmatrix} 0  \\  1 \end{pmatrix}$, which is empty at $t = - \infty$; $\epsilon _{1}$ and $\epsilon _{2}$ are the energies of the two states; and $\epsilon _{12}$ and $\epsilon _{21}$ correspond to the interaction between the two states. According to Ref.~\cite{zener_1932}, the transition probability from $\psi _{+}$ to $\psi _{-}$ will be \\
\begin{equation}
\Gamma \left( \vec{k} \right) \, \sim \, e ^{\frac{- \pi}{2 \hbar \Delta _{\vec{k}}}} ,
\end{equation}

\noindent with $\Delta _{\vec{k}} ^{-1} = \frac{4 \epsilon _{12} ^{2}}{| \frac{d}{d t} \, \left( \epsilon _{1} - \epsilon _{2} \right) |}$. With the transition rate, the excitation density can be estimated through the integral of the probability over the first Brillouin zone, namely, $n _{topo} \sim \int _{1BZ} \, d ^{d} \, \vec{k} \, \, \Gamma \left( \vec{k} \right)$.  

For the Haldane and checkerboard models we consider here, two free-fermion models, the many-body problems can be reduced to one-body problems since the eigenstates of the many-body Hamiltonian must be the Slater determinant of the single-particle states. In addition, with full translation symmetry, there are two good quantum numbers, namely, the two components of the 2D momentum $\vec{k}$. Associated with two sublattices, the Haldane model and checkerboard model therefore become collections of two-level systems (one for each $\vec{k}$), so the LZ theory, which considers a transition of two levels, can be applied. As we mentioned earlier, the particles in the conduction bands and holes in the valence bands generated by the quench process are the topological defects in our models thanks to the analogy between the quantum Hall effect and superfluidity. In the following, we will apply LZ theory to both the Haldane and checkerboard models and compare the results with those of KZ theory.

For the Haldane model near the $K$ (or $K ^{\prime}$) point where the gap closes at critical time, the low-energy Hamiltonian reduces to \\
\begin{equation}
H _{K} \left( \vec{k} \right) = \begin{pmatrix}
m \left( t \right) + \bar{\eta}  &  v _{f} \,  k \, e ^{-i \phi}  \\[0.2cm]
v _{f} \, k \, e ^{i \phi}  & -m \left(  t \right) - \bar{\eta}
\end{pmatrix} ,
\end{equation}
where $m \left( t \right) = \frac{t}{\tau}$, with $\frac{1}{\tau}$ being the quench rate; $\bar{\eta}$ is a constant; $v _{f}$ is the velocity of particles near the $K$ (or $K ^{\prime}$) point; and $\vec{k} = k e
^{-i \phi}$ is the 2D momentum. Note that the linear dependence of $|\vec{k}|$ in the off-diagonal matrix elements captures the property of the linear dispersion near the $K$ (or $K
^{\prime}$) point. As a result, the transition rate of the two energy levels $\Gamma \left( \vec{k} \right) \sim e ^{\frac{-\pi v _{f} ^{2} |\vec{k}| ^{2} \, \tau}{\hbar}}$
gives the topological defect density $n _{topo} \sim \int d \vec{k} \, \, \Gamma \left( \vec{k} \right) \sim \frac{h}{v _{f} ^{2} \tau} \propto \tau ^{-1}$, a power-law relation
with the same power as KZ theory predicts.

For the checkerboard model, setting $\xi ^{\prime} = -\xi ^{\prime \prime} = 0.5$ to simplify the calculation without loss of generality, the Hamiltonian near the gap-closing point $\bar{K} = \left( k _{x}
, k _{y} \right) = \left( \pi, \pi \right)$, after a unitary transformation such that it satisfies the initial condition LZ theory requires that one band is fully occupied and the other is empty, can be expressed as
\\
\begin{equation}
H _{\bar{K}} \left( \vec{k} \right) = \begin{pmatrix} -V  &  \frac{-i}{4} k ^{2} \,  e ^{2i \theta} \\[0.2cm]
             \frac{i}{4} k ^{2} \,  e ^{-2i \theta}  & V
\end{pmatrix} ,
\end{equation}
where $\vec{k} = k \, e ^{i \theta}$ and $V = \frac{t}{\tau}$. In this model, the off-diagonal matrix elements have quadratic dependence on $|\vec{k}|$ due to the quadratic band structure near the band-touching point.
Thus, the transition rate $\Gamma \left( \vec{k}  \right) \sim e ^{\frac{- \pi \tau k ^{4}}{16 \hbar}}$, and $n _{topo} \sim \tau ^{-0.5}$, which also agrees with the prediction of KZ theory.

\section{Quench dynamics with Edges}\label{dynamics}
The presence of edges breaks translation symmetry at least in one direction, and we can no longer reduce the problem to a collection of two-level systems. Instead, we numerically study the slow quench process in the two models by considering strips infinitely long in the $y$ direction with finite width in the $x$ direction with open boundary conditions (OBCs), keeping $k _{y}$ as a good quantum number.

Initially, we prepare the system in the ground state of the initial Hamiltonian $H _{0}$ and consider the half-filling case in which the lower bands are filled. During the slow quench process, we divide the whole process into many time periods such that the Hamiltonian $H \left( t \right)$ barely changes during each period $\Delta t$. From $t$ to $t + \Delta t$, the eigenstate of $H \left( t \right)$ evolves approximately with a phase $e ^{-i \frac{E _{\alpha} \left( t \right) }{\hbar} \Delta t}$, where $E _{\alpha}$ is the corresponding eigenenergy. Taking advantage of the simple evolution of the instantaneous eigenstates, we expand the wave function by the set of the instantaneous eigenstates of $H \left( t \right)$ so that the evolution of the wave function from $t$ to $t + \Delta t$ can be expressed as $\psi _{\beta} \left( t +
\Delta t \right) = \sum\limits_{\alpha} \, e ^{-i E _{\alpha} \left( t \right) \Delta t / \hbar} \, |\phi _{\alpha} \left( t  \right) \rangle \langle \phi _{\alpha} \left( t \right) | \psi _{\beta} \left( t \right) \rangle$, where $\beta$ labels the $\beta$th eigenstate of $H _{0}$ and $| \phi _{\alpha} \left( t \right) \rangle$ and $E _{\alpha}
\left( t \right)$ represent the $\alpha$th eigenstate of $H \left( t \right)$ and the corresponding eigenenergy. At the end of the quench process, we count the contribution of
the initial states to each eigenstate of the final Hamiltonian $H _{f}$.

\section{Numerical results}\label{result}
\subsection{Bulk excitation}\label{bulk}
\begin{figure}[!t]
\centerline{\includegraphics[width=9cm]{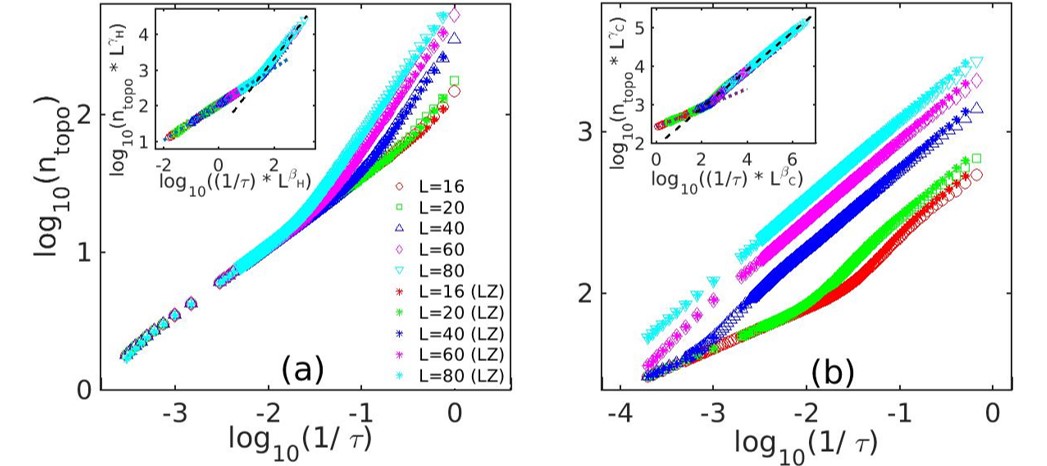}}
\caption{Topological excitation density versus quench rate in a log-log plot for (a) the Haldane model and (b) the checkerboard model. Considering strips infinitely long in the $y$ direction with $L$ sites in the $x$ direction with PBC (no edge), our data and the results from Landau-Zener (LZ) theory are shown for comparison. We perform finite-size scaling on our data such that $n _{topo} ^{\prime} \left( \tau ^{-1} L ^{\beta _{a}} \right) = L ^{\gamma _{a}} n _{topo} \left( \tau ^{-1} \right)$ is size independent in the insets with $\beta _{a}$ and $\gamma _{a}$ being the scaling parameters and $a$ being the model label ($H$ for the Haldane model and $C$ for the checkerboard model.) According to LZ theory, $\tau _{a} ^{-1} \sim k ^{\beta _{a}} \sim L ^{-\beta _{a}}$, with $\beta _{H} = 2$ and $\beta _{C} = 4$. In addition, $n _{topo} \sim L ^{-1}$, meaning $\gamma _{H} = \gamma _{C} = 1$. Our fitting scaling parameters are $\beta _{H} = 1.9332$, $\gamma _{H} = 0.9701$, $\beta _{C} = 4.2266$ and $\gamma _{C} = 1.063$ consistent with the scaling analysis. The black lines are the fitting lines for the powers $\alpha$. In the Haldane model, $\alpha = 1.0000$ in the (relatively) fast quench regime (slope of the dashed line), and $\alpha = 0.5008$ in the slow quench regime (slope of the dotted line). In the checkerboard model, $\alpha = 0.4925$ in the fast quench regime, and $\alpha = 0.2502$ in the other regime. In each model, five systems with $L$ = 16, 20, 40, 60, and 80 are considered. }
\label{HCpbc}
\end{figure}

In order to test the numerical accuracy, we first follow the quench evolution with periodic boundary conditions (PBCs) applied in the $x$ direction, instead of OBCs. In this case there is actually no edge, and we can reduce the problem to two-level systems, as discussed earlier. This allows us to compare our numerical results with those of the LZ theory (taking discrete $k _{x}$ values for finite strips). In Figs.~\ref{HCpbc}(a) and \ref{HCpbc}(b), our data in both models agree with LZ theory very well. In addition, we find that the excitation density $n _{topo}$ and the quench rate $\frac{1}{\tau}$ have power-law relations with different powers $\alpha$ in the (relatively) fast-quench and slow-quench regimes; $\alpha$ in the fast-quench regime agrees with the theoretical value that the KZ and LZ theories predict, but it becomes half of the expected value in the slow-quench regime. In the inset of Fig.~\ref{HCpbc}(a) for the Haldane model, $\alpha$ (slope) at fast quenches is 1.0000 and becomes 0.5008 at slow quenches; in the inset of Fig.~\ref{HCpbc}(b) for the checkerboard model, $\alpha$ is 0.4925 and 0.2502 in the fast- and slow-quench regimes, respectively. The halved powers in the slow-quench regime are due to the finite-size effect. For systems with finite width in the $x$ direction, $k _{x}$ takes discrete values, so that the transition occurs dominantly at $k _{x} = 0$ if we shift the critical point to the origin; at other $k _{x}$ values far from the origin, the dynamics is adiabatic. Therefore, the diabatic dynamics becomes one-dimensional (along $k _{y}$ at $k _{x} = 0$), leading to an exponent half that of the 2D system.

\begin{figure}[!t]
\centerline{\includegraphics[width=9cm]{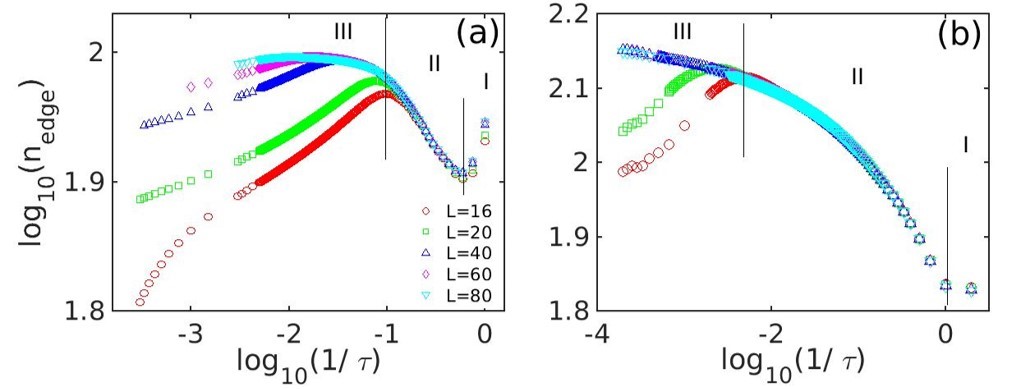}}
\caption{Edge excitation vs quench rate in a log-log plot for (a) the Haldane model with zigzag edges and (b)
the checkerboard model. System sizes: $L$ = 16, 20, 40, 60, and 80 are considered.}
\label{hdck_edge}
\end{figure}

\begin{figure}[!t]
\centerline{\includegraphics[width=8.7cm]{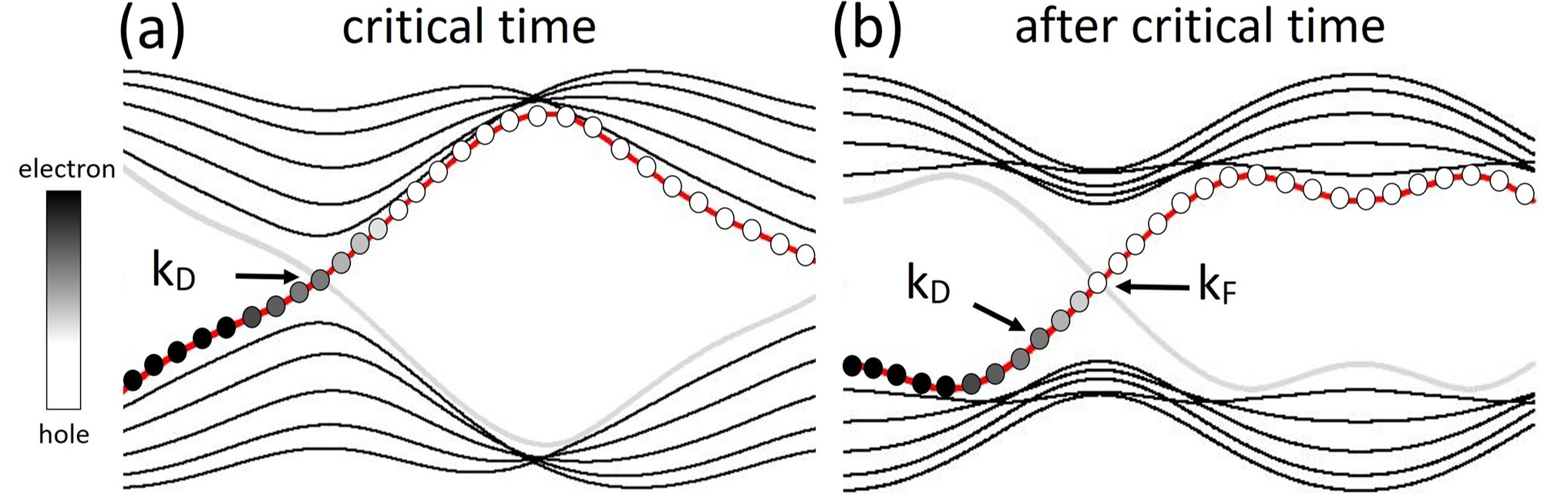}}
\caption{Schematic diagrams for electron occupation of one branch of edge states (a) at the critical time and (b) after the critical time. The light gray curve is the other edge state, which can be ignored in the quasi-adiabatic regime due to the lack of edge-state mixing. $k_{D}$ and $k _{F}$ denote the Fermi energy locations at the critical time and at some moment after the critical time, respectively.}
\label{antikzm}
\end{figure}

\begin{figure}[!t]
\centerline{\includegraphics[width=9cm]{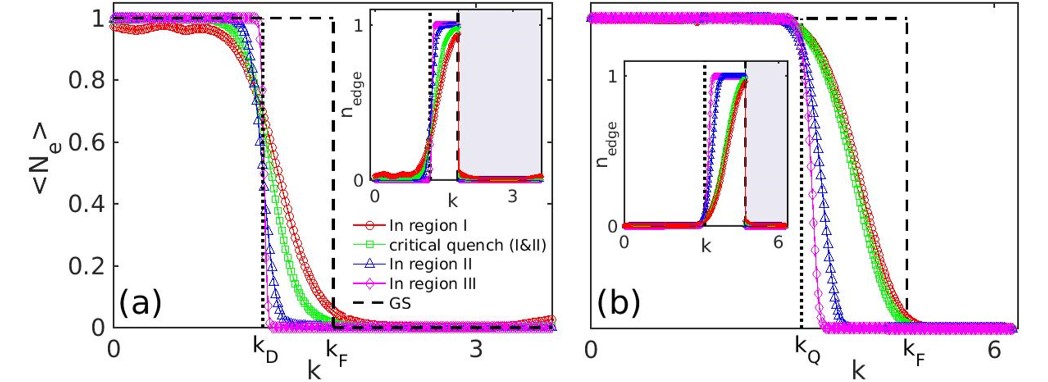}}
\caption{Electron occupation along $k$ with different quench rates for (a) the Haldane model and (b) the checkerboard model. The dotted lines mark $k _{D}$ and $k _{Q}$ (the Fermi level location at the critical time for the Haldane and checkerboard models, respectively), and the dashed curves represent the electron occupation of the edge state in the ground state. The insets show the edge excitation $n _{edge}$ corresponding to each quench rate. The edge excitations in the shaded area are particle excitations, and they are hole excitations otherwise.}
\label{ne_ky}
\end{figure}

\subsection{Edge excitation and the excitation-generation mechanism}\label{edge}
Now turning to the cases with edges, we consider strips with OBCs and zigzag edges for the Haldane model. Since $k _{y}$ is now the only good quantum number, we will simplify the notation by using $k$ for $k _{y}$ in the following. In the presence of edges, a system will have edge states in a topologically non-trivial phase. If the system were in the ground state for the half-filling case, the bands below the Fermi energy $E _{F}$ would all be occupied, and the others would all be empty. Since edge states cross the bulk gap, the electron occupations on edge states in the ground state take the form of a step function $\Theta \left( E _{F} - E \right)$, where $\Theta \left( x \right) = 1$ for $x > 0$ and $\Theta \left( x \right) = 0$ otherwise. Under a quench process, however, the system cannot evolve adiabatically near the critical time and hence cannot stay in the ground state. Instead, excitations will be generated both in the bulk and on the edge states. We can expand the eigenstates of the final Hamiltonian $H _{f}$ through those of the initial Hamiltonian $H _{0}$ (half-filling) at each $k$ as $| \psi \left( k \right) \rangle ^{\left( f \right)} _{\alpha} = \sum _{\beta = 1} ^{L} \omega _{\alpha \beta} \left( k \right) | \psi \left( k \right) \rangle ^{\left( 0 \right)} _{\beta}$, where $| \psi \left( k \right) \rangle ^{\left( f \right)} _{\alpha}$ and $| \psi \left( k \right) \rangle ^{\left( 0 \right)} _{\beta}$ denote the $\alpha$th and $\beta$th eigenstates of $H _{f}$ and $H _{0}$, respectively, with $\omega _{\alpha \beta} \left( k \right) = \prescript{\left( 0 \right)}{\beta}{\langle} \psi \left( k \right) | \psi \left( k \right) \rangle ^{\left( f \right)} _{\alpha}$; $L$ is the width in the $x$ direction (so there are $L$ states.) Assume the states have energies in ascendant order, namely, $E _{1} ^{\left( i \right)} < E _{2} ^{\left( i \right)} < \cdots < E _{L} ^{\left( i \right)}$, with $i = f$ and $0$. The edge states of $H _{f}$ are $| \psi \left( k \right) \rangle ^{\left( f \right)} _{L/2}$ and $| \psi \left( k \right) \rangle ^{\left( f \right)} _{L/2 + 1}$, and the (total) edge excitations $n _{edge}$ can  be defined as \\
\begin{equation}
\begin{aligned}
n _{edge} =  \sum ^{L/2 + 1} _{\alpha = L/2} & \int d k \left[ \right. \sum ^{L/2} _{\beta = 1} | \omega _{\alpha \beta}  \left( k \right) | ^{2} \Theta \left[ \right. E _{\alpha} ^{\left( f \right)} \left( k \right) - E _{F} ^{\left( f \right)} \left.\right] \\[0.2cm]
+ & \sum ^{L} _{\beta = L/2 + 1} |\omega _{\alpha \beta} \left( k \right) | ^{2} \Theta \left[ \right. E _{F} ^{\left( f \right)} - E _{\alpha} ^{\left( f \right)} \left( k \right) \left. \right] \left. \right],
\end{aligned}
\end{equation}
the deviation of the electron occupations on edge states from their ground-state occupations at the end of the quench.  In Figs.~\ref{hdck_edge}(a) and \ref{hdck_edge}(b), we investigate $n _{edge}$ with various quench rates and find distinct behaviors in three regimes of the quench rate. In regime I, $n _{edge}$ decays as the quench rate decreases, consistent with the physical intuition. In regime II, unexpectedly, $n _{edge}$ increases as the quench rate decreases, displaying an anti-KZ behavior. In regime III, it seems to obey the KZ mechanism again.

To understand the behaviors of $n _{edge}$, especially the anti-KZ behavior, we examine the formation and evolution of the edge states, starting with the moment when the band gap closes and edge states just begin to form in the Haldane model. At the critical time, conduction and valence bands strongly mix near the band-gap-closing point $k _{D}$ and form edge states.
Except for the slowest quench rate, which we will comment on later, we can ignore the coupling between states on opposite edges and focus on one branch of the edge states, pretending the other edge does not exist. We can schematically express the branch of the edge states of interest as $\psi _{edge} \left( k \right) = a \left( k \right) \psi _{c} \left( k \right) + b \left( k \right) \psi _{v} \left( k \right)$, where $\psi _{c}$ and $\psi _{v}$ denote the contributions from the conduction and valence bands before the gap closing, respectively, and $a \left( k \right)$ and $b \left( k \right)$ represent the corresponding weights, depending on $k$ but not the quench rate, with $a ^{2} \left( k \right) + b ^{2} \left( k \right) = 1$.
As illustrated in Fig.~\ref{antikzm}(a), edge states with $k < k_{D}$ are predominantly from valence bands, while those with $k > k_{D}$ are predominantly from conduction bands. This feature is expected to persist as the gap opens up, as illustrated in Fig.~\ref{antikzm}(b).
We now introduce a regime of ``quasiadiabatic'' time evolution after the gap opens; namely, the quench rate is small enough that there is no transition between edge states and bulk states but fast enough that edge states remain at the {\em same} edge. That is, an edge state at one edge does not evolve into the other at the opposite edge with the same $k$ when their energies cross. For sufficiently wide strips such a regime is guaranteed to exist as the coupling between the edges is exponentially suppressed. In this regime the occupation number of an edge state is well approximated by $b ^{2} \left( k \right)$ at the point of gap closing, which is close to 1 for $k < k_{D}$ and close to zero otherwise. With the gap opening up, however, the Fermi wave vector $k_{F}$ increases [see Fig.~\ref{antikzm}(b)], resulting in significant numbers of hole-like edge excitations for $k_{D} < k < k_F$. On the opposite edge we expect equal numbers of particlelike excitations. Increasing the quench rate induces relaxations of the edge excitations into the bulk, giving rise to the anti-KZ behavior. Further increasing the quench rate, on the other hand, induces additional edge excitations, especially outside the range $k_{D} < k < k_{F}$. This brings us back to the usual KZ behavior. Figures~\ref{ne_ky}(a) and \ref{ne_ky}(b) show the electron occupation distribution of one edge state along $k$ in different quench regimes, and the insets show the dominance of hole excitations on this edge state, supporting our argument.

Finally, the KZ behavior in regime III is due to essentially true adiabatic evolution in which edge states on opposite edges are mixed.
This is
a finite-size effect due to the exponentially small coupling between the edges. Consistent with this understanding, we find it is more prominent in small systems, as shown in Fig.~\ref{hdck_edge}. In the thermodynamic limit, without the particle leaking from edges, the edge excitation numbers will eventually saturate in the slow quench, instead of going to zero as we observed at finite strip width.

\section{Conclusion}\label{conclusion}
We numerically studied the dynamics of the Haldane model and the checkerboard model in slow-quench processes through topological quantum phase transitions with $\Delta C = 1$ and $2$, respectively. We showed the agreement of the power-law relations of the bulk excitation and the quench rate with the predictions of the KZ and LZ theories. In addition, an anti-KZ behavior of the edge excitation was found in both models. We provided a physical picture for this counterintuitive feature which originates from the unrelaxable nature of excitations on edge states since the edge states form.

{\it Acknowledgments.}
This work was supported by National Science Foundation Grant No. DMR-1442366 and was performed at the National High Magnetic Field Laboratory, which is supported by National Science Foundation Cooperative Agreement No. DMR-1644779 and the state of Florida.

\bibliography{ref}

\end{document}